\documentclass[useAMS,usenatbib]{mn2e}

\usepackage{amssymb}

\input{epsf}
\input psfig.sty

\begin{document}

\title[Angular momentum$-$Large-scale structure alignments]{Angular momentum$-$Large-scale structure alignments in $\Lambda$CDM models and the SDSS}
\author[D. J. Paz, F. A. Stasyszyn, N. D. Padilla]
{Dante J. Paz$^1$, Federico Stasyszyn$^{1,2}$, Nelson D. Padilla $^{3}$ \\
$^1$ Instituto de Astronom\'{\i}a Te\'{o}rica y Experimental IATE
 (UNC-CONICET),\\ Observatorio Astron\'{o}mico C\'{o}rdoba, Francisco N. Laprida 922, C\'{o}rdoba, Argentina.\\
$^2$ Max-Planck-Institut fuer Astrophysik MPA, Karl-Schwarzschild-Str. 1, Garching, Germany.\\
$^3$ Departamento de Astronom\'{\i}a y Astrof\'{\i}sica, Pontificia 
Universidad Cat\'olica de Chile,\\ Vicu\~na Mackenna 4860, Santiago 22, Chile.\\
}

\date{Accepted 25 June 2008. Received 24 June 2008; in original form 25 April 2008}
\maketitle

\begin{abstract}
We study the alignments between the angular momentum of individual
objects and the large-scale structure in cosmological numerical simulations
and real data from the Sloan Digital Sky Survey, Data Release 6. 
To this end we measure anisotropies
in the two point cross-correlation function around simulated halos and observed galaxies,
studying separately the 1- and 2-halo regimes.
The alignment of the angular momentum of dark-matter haloes in $\Lambda$CDM simulations 
is found to be dependent on scale and halo mass.  At large distances (2-halo regime), 
the spins of high mass haloes
are preferentially oriented in the direction perpendicular to the distribution of matter;
lower mass systems show a weaker trend that may even reverse to show an angular momentum
in the plane of the matter distribution.
In the 1-halo term regime, the angular momentum is aligned
in the direction perpendicular to the matter distribution; the effect is stronger 
than for the 1-halo term and increases for higher mass systems.

On the observational side, we focus our study on galaxies in the 
Sloan Digital Sky Survey, Data Release 6 (SDSS-DR6) with elongated apparent shapes,
and study alignments with respect to the major semi-axis.
We study five samples of edge-on galaxies, the full SDSS-DR6 edge-on sample, bright galaxies, faint galaxies,
red galaxies, and blue galaxies (the latter two 
consisting mainly of ellipticals and spirals respectively). 
Using the 2-halo term of the projected correlation function, we find an
excess of structure
in the direction of the major semi-axis for all samples; the red sample 
shows the highest alignment ($2.7\pm0.8\%$) and indicates that the angular momentum
of flattened spheroidals tends to be perpendicular to the large-scale structure.
These results are in qualitative agreement with the
numerical simulation results indicating that the angular momentum of galaxies could be built
up as in the Tidal Torque scenario.  The 1-halo term only shows a significant alignment for
blue spirals ($1.0\pm0.4\%$), 
consistent with the 1-halo results from the simulation but with a lower amplitude.
This could indicate that even though the structure traced by galaxies is adequate to study large-scale
structure alignments, this would not be the case for the inner structure
of low mass haloes, $M\leq 10^{13}h^{-1}M_{\odot}$,
an effect apparently more important around red $g-r>0.7$ galaxies.
\end{abstract}


\section{Introduction}

In the current paradigm of large-scale structure formation in the Universe it is well established
that the angular momentum of dark-matter (DM) halos is determined early in the history of the Universe
by the gravitational
interaction between the quadrupole of the collapsing proto-halo region and the surrounding matter.
This was first formulated for the 
hierarchical theory of structure formation by \citet{White1}, \citet{Doro1}, and \citet{Peebles1}, and it is claimed that
this is a natural consequence of the perturbative treatment of the gravitational instability
scenario (\citet{Porciani1}). 
Several inherent properties to the tidal torque process (Tidal Torque Theory, TTT)
have been noticed and studied in the literature. For instance
it is expected from TTT,
and also supported to some degree by N-body simulations, that  the angular 
momentum starts growing linearly with time \citep{Catelan2,Peebles1,White1}, and that
later on this process looses efficiency 
after the halo turn-around. This is due to the reduction of halo inertia by collapse, and to the
continued expansion of the neighboring matter responsible for the tidal torque. 

There are several analytic studies of the TTT scenario \citep{White1,Doro1,Peebles1,Catelan3},
as well as semi-analytic approaches \citep{Catelan1}, and fully non-linear work using 
cosmological simulations 
\citep{Barnes1,Suger1,Porciani1,Porciani2,Peirani1,Nagashima1}. The agreement is that the resulting halo 
population is poorly rotationally supported; the velocity dispersion at virialisation is found to be 
the most important mechanism for halo equilibrium. 
As a result, the distribution of spin parameters\footnote{
The dimensionless spin parameter as defined by \citet{Peebles1} is $\lambda=L\sqrt{|E|}/GM^{5/2}$, 
where $E$ is the total internal energy of the halo and $L$ and $M$ are its angular momentum and mass 
respectively. This number can be understood
as a measure of the importance of rotational velocity
vs. velocity dispersion as the dominant support mechanism of a halo} 
is log-normal which peaks at quite low values, $\lambda_{med} \simeq 0.035$ \citep{Bullock1,Gardner1, 
Onghia1, Maccio1, Bett1}. Unfortunately it is not possible to obtain a direct measurement
of spin parameters
in actual galaxies. Despite this,  
\citet{Hernandez1} used indirect methods to
derive distributions of galaxy spin parameters 
consistent with
theoretical predictions.

From an observational point of view, it can be easier in some cases 
to determine the direction of the angular 
momentum of a given galaxy rather than its amplitude.  There has been a recent increase in the 
interest concerning the possibility of galaxy alignments
with the large scale structure. Such studies are important to test the already well developed 
TTT, and to asses the possibility that such correlations affect the results of cosmic shear measurements.
Regarding the former, TTT produces noisy predictions due to the non-linear nature 
of the accretion processes of clumpy matter \citep{Vitvitska1,Porciani1}, on scales below the 
coherent scale of flows predicted by linear theory. 
The second tests for the effects on cosmic shear studies used to analyse weak lensing, may answer
whether there is an intrinsic alignment in the structure traced by galaxies that could result 
in a spurious contribution to the shear power spectrum \citep{Hirata1,Bridle1,Takada1,King1,King2,Heymans1,Heymans2,
Brown1,Catelan4,Crittenden1,Croft1,Jing1}.
Several authors have searched for galaxy alignments in the super-galactic plane \citep{Vauco1},
extending out to several tens of megaparsecs from our Galaxy, and more recently on large 
galaxy surveys \citep{Hirata1,Hirata2,Lee1,Lee2,Mandel1}.  
\citet{Lee1} claim a direct evidence of an alignment between galaxy
position angle in 2MRs and the tidal field deduced 
from galaxy positions.

 An interesting approach to the analysis of galaxy alignments is to use the radial direction to
the centres of voids
as characteristic directions in the large scale structure. It is well known that 
as they evolve, large voids tend to be rounder, a behaviour which is opposite to that shown by
bound structure.
Negative density perturbations in the initial fluctuation field
that will later become voids,
are characterised by a decreasing asphericity as the expansion of the Universe proceeds,
as is shown by \citet{Sheth1} (and references therein). 
Taking advantage of this particular characteristic of voids, it is possible to study the
whether galaxy or halo angular momenta are aligned with the void shells just by
considering the angle with respect to the direction to the void centre.
This is the approach used by \citet{Trujillo1} applied to the SDSS and 2dFGRS
galaxy redshift surveys, and by \citet{Brunino1}, \citet{Cuesta1} and \citet{Patiri1} to numerical simulations. 
\citet{Trujillo1} find with a high level of confidence that spiral galaxies located on the shells of
the largest cosmic voids have rotation axes that lie preferentially on the void surface, in agreement with the simulation
results of \citet{Brunino1,Cuesta1,Patiri1}.
Other more general methods that also take into account alignments with large scale filaments in
numerical simulations show similar results
\citep{Aragon1,Hahn1,Hahn2}.

In this paper we present results from a study of alignments between individual
objects and their surrounding structure in quantitative concordance with TTT predictions and
previous observational results. We use a novel approach to study alignments
with the large-scale structure using as a reference the galaxy/dark matter halo angular momentum.
Using numerical simulations we study alignments predictions in a Cold Dark Matter model with a cosmological constant
($\Lambda$CDM), and compare such predictions with the results obtained from the Sloan Digital Sky Survey,
Data Release 6 (SDSS-DR6, Adelman-McCarthy et al., 2007). To this end we use the galaxy
two point correlation function applied both, in its three-dimensional definition to
the full numerical simulation and in its projected version to mimic the observational limitations.  
The latter allows us to determine that we should be able to detect the expected low signal alignments, 
even in small galaxy samples. 
This allows us to measure the alignment of galaxies and halos over wide ranges in mass and separation,
and also to quantify the dependence of this alignment on halo mass, galaxy colour and luminosity.

This paper is organised as follows.  Section 2 presents the numerical simulation and the statistical
tools we will use in this work; it also presents an analysis of the impact of the observational biases on the
detection of alignments.  Section 3 contains the analysis of observational samples of galaxies
from the Sloan Digital Sky Survey.  The comparison between models and observations
as well as the discussion of the results is presented in Section 4, and
Section 5 contains the final conclusions from this work.

\section{Anisotropies in the 3D Correlation function}

In this section we analyse the three-dimensional two-point correlation function, 
using cluster-particle pairs, 
measured using $\Lambda$CDM numerical simulations, taking into 
account the orientation with respect to the centre halo angular momentum. 
To this end, we have run two periodic simulations of $500$ and $60$ $h^{-1 }$Mpc 
computational box sizes, for a flat low-density Universe, 
with a matter density $\Omega_{\rm m}=1-\Omega_{\Lambda}=0.3$, Hubble constant 
$H_{\circ}=70$ km s$^{-1}$ Mpc$^{-1}$, and normalisation 
parameter $\sigma_{8}=0.8$. The particle resolutions are $m_{\rm p}\ge 
7.2\times 10^{10}\,h^{-1}\,M_{\odot}$ for the large simulation, and 
$m_{\rm p}\ge 1.2\times 10^{8}\,h^{-1}\,M_{\odot}$ for the small one. 
The identification of particle clumps has been carried out by 
means of a standard friends-of-friends algorithm with a percolation 
length given by $l=0.17$ $\bar{\nu}^{-1/3}$, where $\bar{\nu}$ is the mean number density 
of DM particles. 
Both simulations have been performed using the first version of the GADGET 
code developed by Springel et al. (2001). We consider halos with 
at least $30$ and $100$ particles in the large and small simulations, respectively, 
in order to ensure an accurate measurement of the
halo angular momentum.

\begin{figure*}
\begin{picture}(430,250)
\put(-40,0){\psfig{file=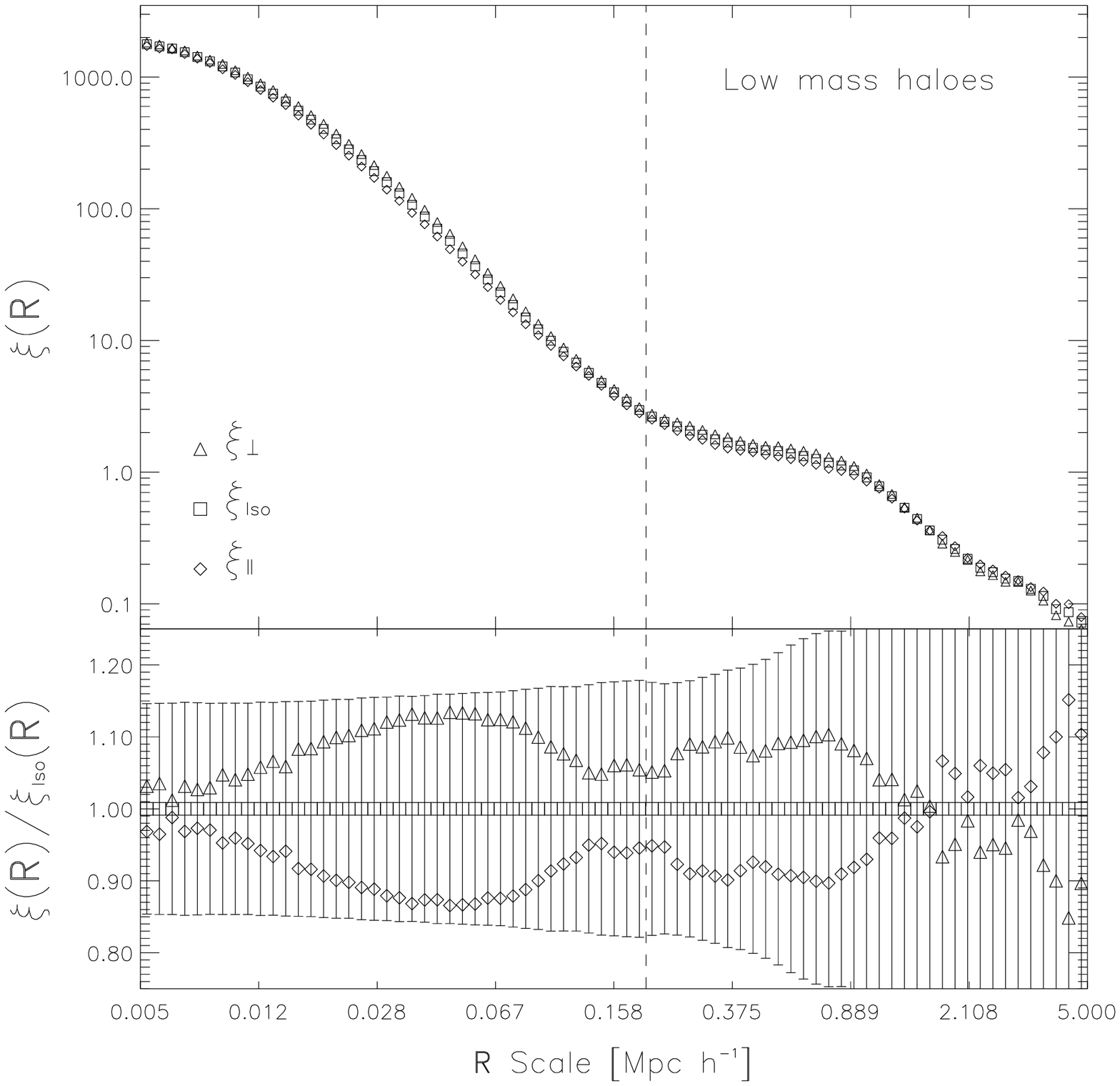,width=9.cm}}
\put(210,0){\psfig{file=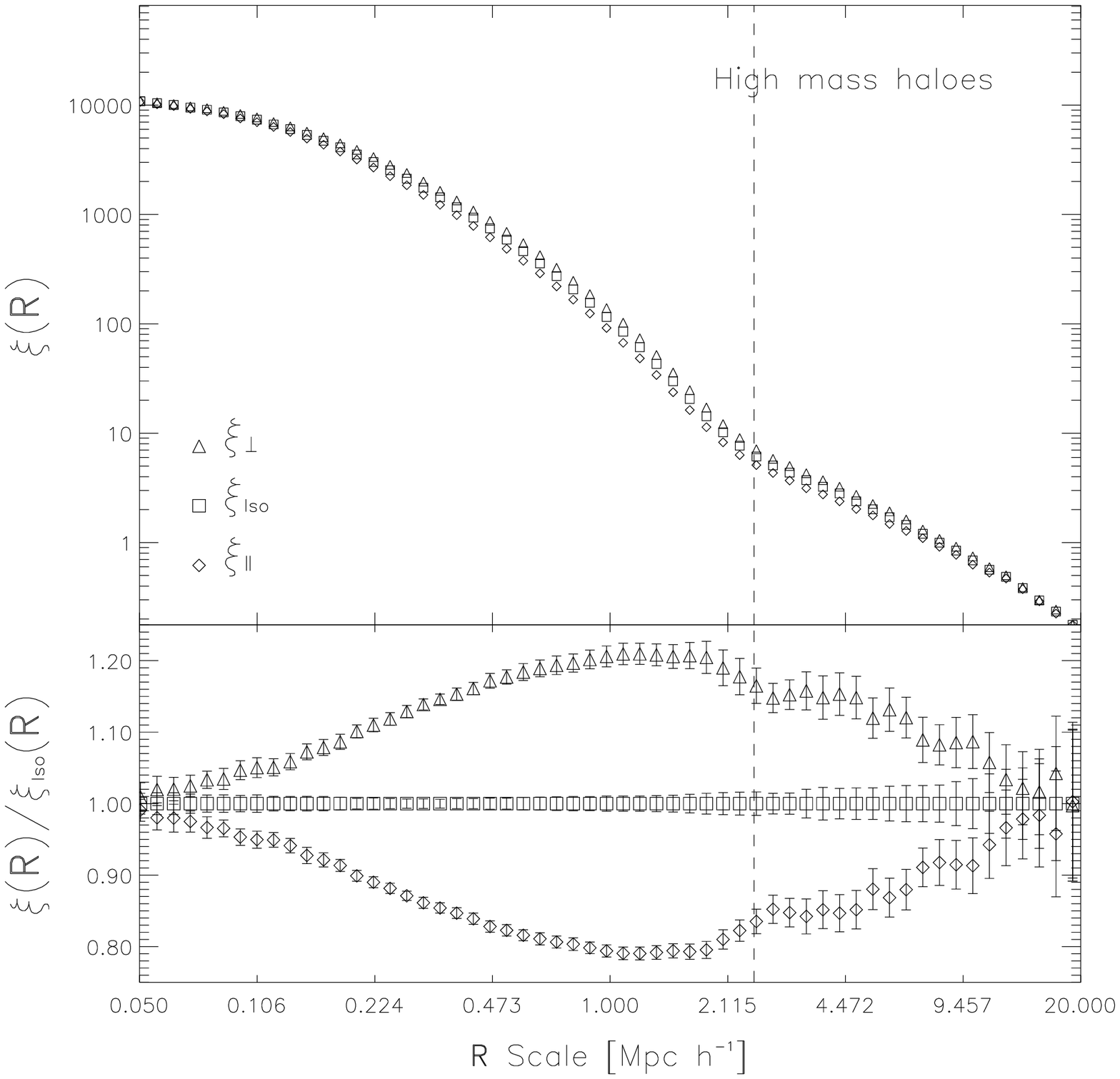,width=9.cm}}
\end{picture}
       \caption{
Left panel:
Spatial Halo$-$Dark-matter
correlation function of low mass haloes in the small numerical simulation.  
The triangles correspond to the correlation function between haloes and neighbors in the direction 
perpendicular to the angular momentum of the centre haloes.  The diamonds show the results when using
tracers along the direction of the angular momentum, and the squares the results when all neighbors
are taken into account.  The lower left panel shows the ratio between the correlation functions shown in the 
top-left panel
and the results using all neighbors.
Right panel: same as left panel, for a high mass DM halo subsample from the large simulation.
}
       \label{fig:3D}
\end{figure*}

The spatial halo-particle cross-correlation function $\xi(r)$ which we will use in this section,
measures the excess probability
with respect to a random distribution, that a DM particle
will reside at a distance $r$
away from a given halo centre, in the volume element $dV$,
\begin{equation}
dP=\bar{\nu}\left[ 1+\xi(r)\right] dV.  \label{definicion}
\end{equation}
A standard method to measure $\xi(r)$ in a simulation box 
is to count pairs at a given distance bin, and then normalise this
number by the expected number of pairs for a constant number density of objects. 
In our measurements we perform these counts stacking pairs 
depending on the angle between their position relative to the centre halo and its angular momentum. 
We consider tree cases, (i) using all pairs regardless of this angle
(spherical shell volumes, isotropic case), 
(ii) using all pairs subtending an angle from the angular momentum 
less than a given threshold $\theta_1$ (parallel case), 
and (iii) adding pairs at lower inclinations than a limit angle 
$\theta_2$ from the perpendicular plane to the angular momentum (perpendicular case).
We choose the threshold angles, for the parallel and 
perpendicular cases mentioned above ($\theta_1$ and $\theta_2$, respectively), so
that the volumes in each case are the same.
This can be achieved by setting, $\sin(\theta_2)=1-\cos(\theta_1)=\chi$, and choosing a value for the
threshold parameter $\chi$. Selecting $\chi\leq0.5$ implies angles $\theta_1 \leq 60^o$ 
and $\theta_2 \leq 30^o$.  Throughout this paper we use $\chi=0.5$.

Errors in the correlation function are estimated using the jacknife technique, with a total of $50$ subsamples 
for the numerical simulation  ($20$ jacknifes for the SDSS data analysis). We find that our errors are
stable for this number of jacknife subsamples (we tested from $20$ to $100$) and 
does not produce underestimations.  
In the case of quotients between correlation functions, we apply the jacknife technique to the 
actual measurement of the ratio. We adopt the jacknife technique since it has been reported to provide 
similar error estimates to those obtained from independent samples (see for instance, Padilla et al., 2004).

\subsection{Simulation results}\label{simresult}

The analysis of alignments of structure with the angular momentum of DM haloes presents
a first difficulty in that in order to obtain a reliable and stable direction for the halo spin, it is
often necessary to use a large number of particles per halo.  We find that using $30$ particles
provides reasonable results, but this imposes a rather high lower limit to halo masses of $2.2\times
10^{12}\,h^{-1}\,M_{\odot}$ for the large simulation, and $3.6\times 10^{9}\,h^{-1}\,M_{\odot}$ for 
the smaller one.  We analised the distribution of angular momentum parameters and find that
the reported high $\lambda$ tail
(Bett et al., 2007) due to low number of particles per halo, is not present
in any of our samples of DM haloes.
We notice that the clustering signal in
the small simulation on large scales may be affected by the small box size.

From this point on, we will centre most of our analysis on samples of DM haloes with different masses.  
From the small
simulation we draw four DM halo subsamples separating haloes of different mass starting at
a minimum of 
$100$ particles or $M\geq 1.2 \times 10^{10}\,h^{-1}\,M_{\odot}$ (the lowest mass sample considered includes
haloes of up to 300 particles each; we do not use haloes of lower masses even though we are still above
the minimum number of particles we consider safe to consider).
Nine samples are drawn from the large simulation and contain DM haloes with more
than $30$ particles.  The subsample with the most massive haloes starts at $1000$ particle
members corresponding to
$M\geq 7.2\times 10^{13}\,h^{-1}\,M_{\odot}$.

\begin{figure}
       \epsfxsize=0.5\textwidth 
       \vspace{0.5cm}\hspace*{-0.5cm} \centerline{\epsffile{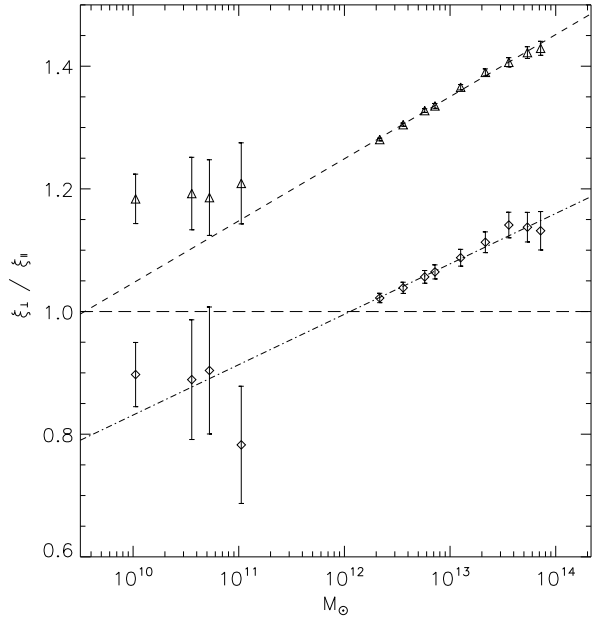}}
       \caption{ Ratios between correlation functions for neighbors in the 
                 directions perpendicular and parallel to the halo angular 
		 momentum. Diamonds (triangles) correspond to the ratio over the 2-halo (1-halo) term scales.  
               }
       \label{fig:rat3d}
\end{figure}

The left panels of Figure \ref{fig:3D} show the cross-correlation function between low mass
DM haloes and DM particles 
(top-left) in the directions parallel and perpendicular to their angular momenta,
and in all directions (diamonds, triangles and squares, respectively).  The lower panel shows
the ratios between these three estimates and the correlation function considering all neighbors; 
their relative amplitudes help identify alignments in either direction.  
As can be seen, the relative amplitudes of the correlation function when using neighbors
in the directions parallel and perpendicular to the angular momentum of the DM
halo are different; for small separations $R<1$h$^{-1}$Mpc, there is an excess of correlation
in the direction perpendicular to the angular momentum of up to a $20\%$ effect (between
correlations parallel and perpendicular to the angular momentum, lower panel),
whereas for larger separations 
the alignment diminishes and shows a 
tendency to reverse (although, as can be seen in the figure, errors are larger at these 
separations); we come back to this later in this section where we use a higher signal estimator 
for alignments. 

When using higher mass DM haloes, the effect is also present with an even 
higher difference and statistical significance between 
the structure in the direction of the angular momentum and perpendicular to it (Right panels 
of Figure \ref{fig:3D}). In this case, neighbors in the direction
perpendicular to the angular momentum show more correlation, and there is
a clear tendency for this alignment to decrease with separation, which goes from 
almost a $50\%$ effect on relative amplitudes (lower panel) at small separations 
to about $<5\%$ at large separations.  Note that the alignment does not change
at large separations as seemed to be the case for low mass haloes, even at
$R=20$h$^{-1}$Mpc compared to $R=5$h$^{-1}$Mpc analised in the low mass halo sample.

Notice that the halo-particle cross-correlation functions in Figure \ref{fig:3D} show
the characteristic 1- and 2-halo terms reported in several previous measurements
of both observational (e.g. Zehavi et al., 2004, Cooray, 2005)  and numerical
simulations (Zheng et al., 2004); these two regimes are separated in the figure by the 
vertical dashed lines.
This separation is obtained by measuring the scale at which the correlation function is
seen to shift between these two regimes (there is a local minimum in the first derivative of
the correlation function).
It can be seen that the alignment signal shows a clear transition from
the 1- to the 2-halo regime; in particular, low mass haloes
show indications of a different angular momentum$-$large-scale alignment behaviour in the 2-halo regime, suggesting 
an important relationship between alignment and halo term.
This can be interpreted as different physical mechanisms producing the alignment in the
1- and 2-halo terms.

We make quantitative estimates of the alignments for the 1- and 2-halo terms separately from
the cross-correlation functions shown in Figure \ref{fig:3D} as well as from the
other samples of DM haloes with intermediate 
masses (for a total of 13 samples of varying halo masses).
We only consider pairs that are separated by a distance at least as large as the 
transition between regimes
for the 2-halo term, and at shorter separations for the 1-halo calculations.

In order to avoid effects of covariance between the correlation function bins to some extent, 
we make global 1- and 2-halo term estimates of the ratios between the correlation functions
in the directions perpendicular and parallel to the angular momentum of haloes, using all the
halo-particle pairs
separated by distances within the 1- and 2-halo ranges (i.e. we do not use the narrow bins
in $\log_{10}(r)$ used to calculate the cross-correlation functions).  
We acknowledge that traces of covariance will still
be present between the 1- and 2-halo term ratios.
We present these results in Figure \ref{fig:rat3d}, where triangles show the 1-halo term ratio
between correlation functions for neighbors in the directions 
perpendicular and parallel to the halo angular momentum;
Diamonds correspond to the 2-halo term.  The horizontal long dashed line at $\xi_\parallel/\xi_\perp = 1$
corresponds to the isotropic case.
The data-points at masses lower than $10^{12}h^{-1}M_{\odot}$ correspond to the small numerical
simulation, and show larger errors due to both, the smaller simulated volume, and the lower amplitude
of clustering characterising low mass haloes.

As can be seen, the 1-halo term always shows more structure in the direction
perpendicular to the angular momentum, and more so for larger masses, whereas the 2-halo term
shows a similar behaviour with the possible exception of low mass haloes, whose angular momenta tend 
to point in the direction of the large-scale structure (although with a low statistical significance).
The alignments are detected at about a $5-\sigma$ level for the higher mass halo subsamples drawn
from the large simulation. As in the 2-halo regime, low mass haloes show considerably higher errors. 

In the case of low mass halos ($M\simeq 10^{10}h^{-1}M_{\odot}$) in the 2-halo term regime (diamonds in Figure
\ref{fig:rat3d}), the alignment signal is the opposite to that of higher mass haloes (a $10\pm5\%$ effect). 
This anti-alignment signal fits in well with
the overall tendency for this quotient $\xi_\parallel/\xi_\perp$ to increase with mass, which can
be fit by a log-linear
relation which crosses the unit value at $M\simeq10^{12}h^{-1}M_{\odot}$. This mass limit suggest a change
in the alignment signal in qualitative agreement (regarding the alignment direction) with results 
on simulations by \citet{Hahn1}, with masses above and below the typical
mass scale for collapse $M_*=5.5\times10^{12.5}h^{-1}M_{\odot}$ at $z=0$. The best-fit log-linear
relation for the 2-halo term results is shown in Figure 
\ref{fig:rat3d} (dash dotted line), and corresponds to,
\begin{equation}
\xi_\parallel/\xi_\perp = (0.95\pm0.01) + (0.082\pm0.008) *[log_{10}(M)-11.5]\,\,
\end{equation}
For the 1-halo term, the dependence of alignment on mass is well adjusted by (dashed line on Figure \ref{fig:rat3d}),
\begin{equation}
\xi_\parallel/\xi_\perp = (0.95\pm0.01) + (0.082\pm0.008) *[log_{10}(M)-11.5]\,\,
\end{equation}

The 2-halo term effect can be understood within the framework of TTT, specially
for the high mass haloes, which readily show that their angular momentum is perpendicular
to the present-day mass distribution.  It seems that for high enough mass haloes, the 
local large-scale structure still describes the original tidal field.  In the case of low
mass haloes, the alignment seems to reverse.  This result can be 
understood by an intuitive non-linear effect beyond the abilities of TTT, for example
if these haloes
have formed recently in filaments, from the merger of material arriving from voids surrounding
these elongated structures.

The 1-halo term results indicate that regardless of halo mass, the structure within
dark-matter haloes is preferentially aligned with the plane perpendicular to the angular
momentum, with a higher statistical significance and signal than the 2-halo term. 

\begin{figure*}
\begin{picture}(430,250)
\put(-35,0){\psfig{file=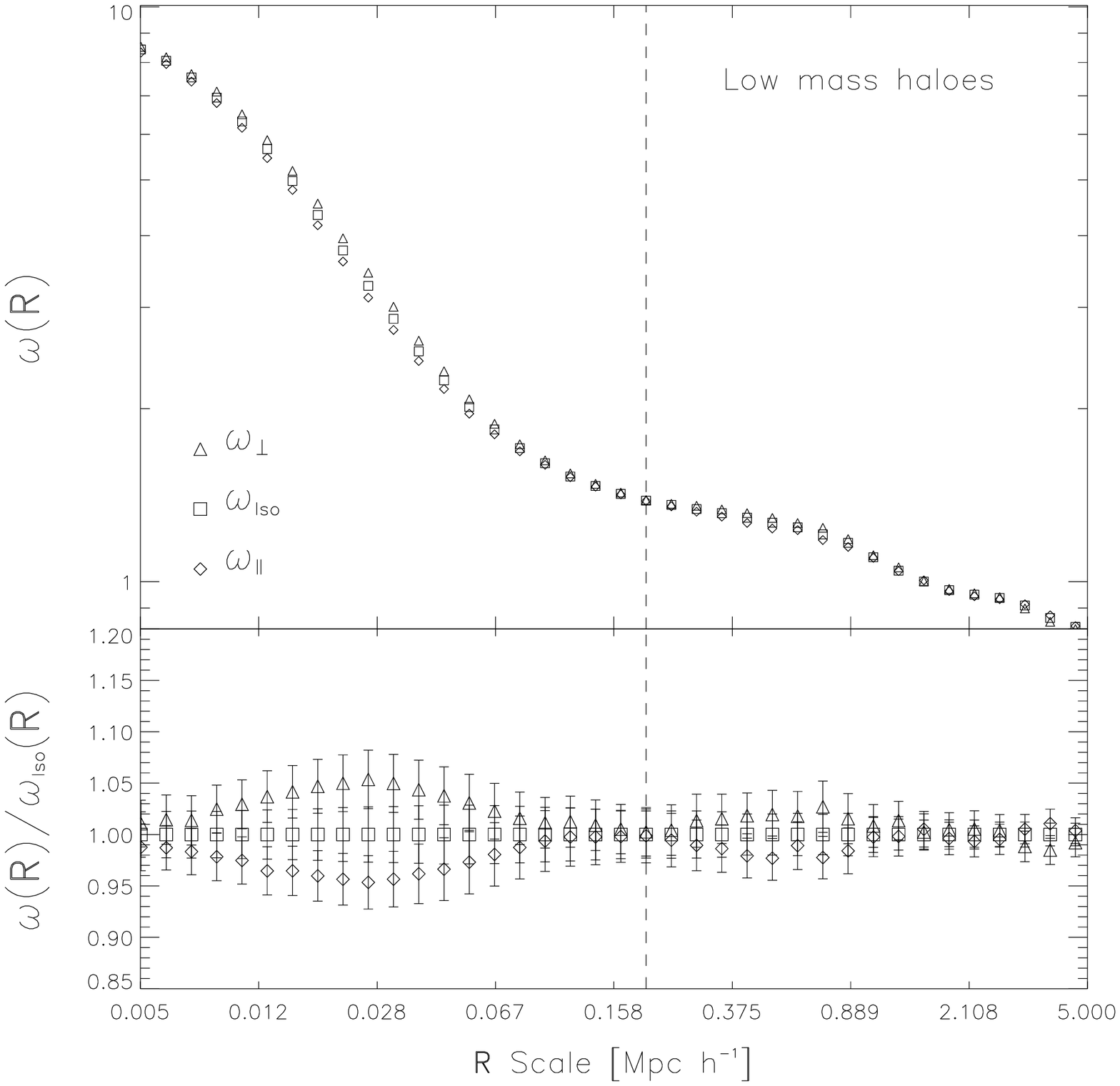,width=9.cm}}
\put(215,0){\psfig{file=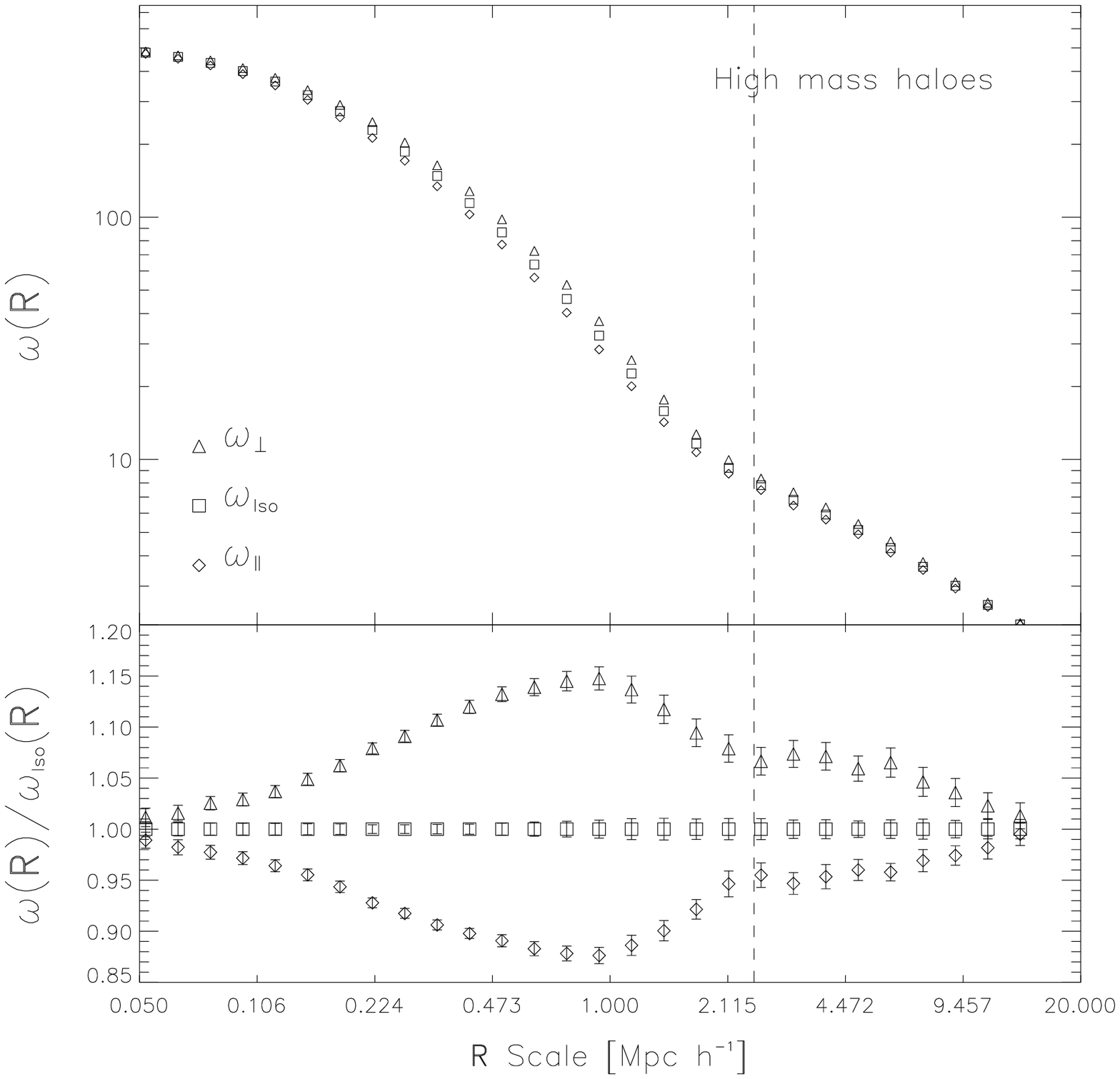,width=9.cm}}
\end{picture}
       \caption{
Left panels: projected 
correlation functions around "edge-on" low mass haloes in the small numerical simulation (top-left), 
in the directions
perpendicular and parallel to the angular momentum of centre haloes (triangles and
diamonds symbols, respectively).  Ratios between these projected correlation functions are shown
in the bottom-left panel.  Right panels: same as left panels for high mass haloes in the large
numerical simulation.
}
       \label{fig:projsim}
\end{figure*}

\begin{figure}
\begin{picture}(350,420)
\put(15,240){\psfig{file=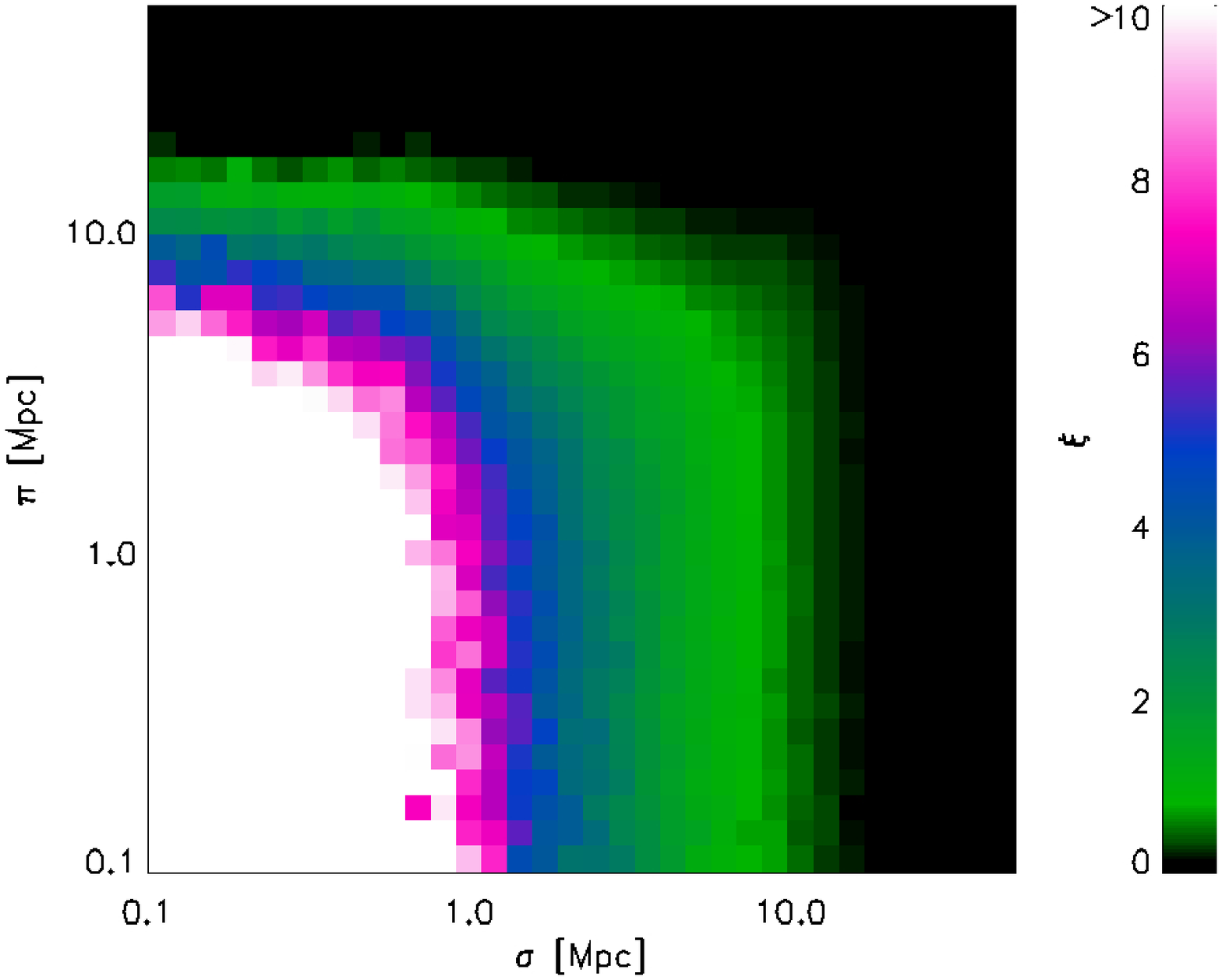,width=8.cm}}
\put(0,0){\psfig{file=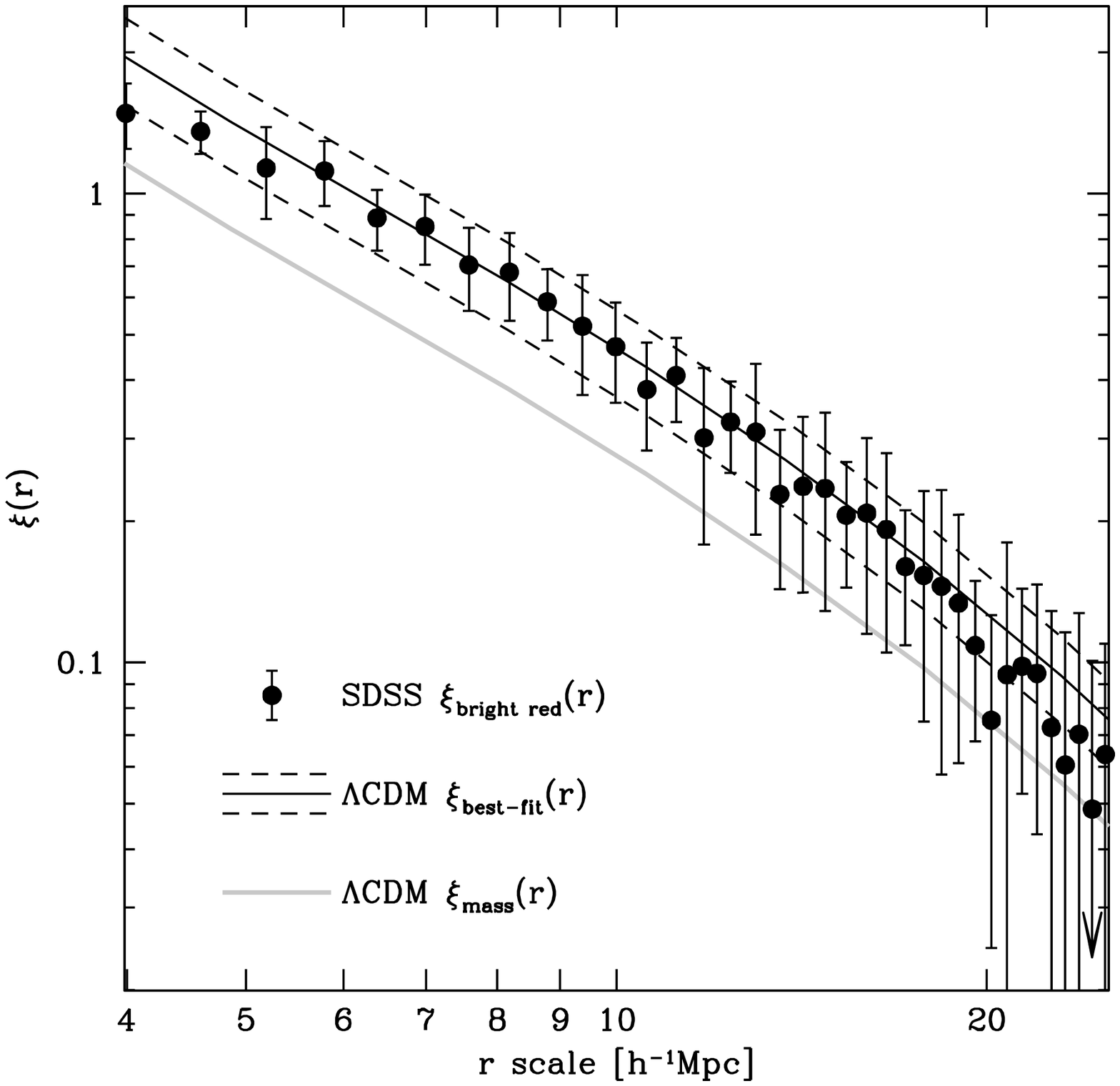,width=8.5cm}}
\end{picture}
       \caption{
Top panel: 
Correlation function around "edge-on" red spiral galaxies in the SDSS-DR6 (sample O3), in the directions
parallel and perpendicular to the line of sight.  The colour indicates the amplitude of the correlation
function (see the figure key).  Bottom panel: real-space correlation function for the same sample
of galaxies, obtained as explained in the text (symbols with errorbars), and the best-fit $\Lambda$CDM
real-space correlation function (black lines).  The dashed lines show the range of correlation function
amplitudes allowed by the errors in the data.  The gray thick line shows the mass real-space
correlation function in the $\Lambda$CDM model.
}
       \label{fig:xir}
\end{figure}

\subsection{Simulating observational biases in the numerical simulation}

It can be difficult to estimate the angular momentum of real galaxies.  In the case of the
SDSS-DR6, which we use in this work, the only information
that can allow us to obtain an estimate of the direction of the angular momentum 
is the photometry of galaxy images, specially for spiral galaxies but also for ellipticals.  
We assume that spiral discs or flattened spheroids
are perpendicular to the angular momentum of galaxies and 
therefore, edge-on galaxies (either disks or spheroidals)
will be more likely to have their angular momentum on the plane of the sky.
Then, one can calculate the cross-correlation function for neighbors in the projected direction parallel
and perpendicular to the direction of the angular momentum of the centre galaxies, 
in redshift slices around them, assuming all these neighbors to be at the same distance
than the centre galaxy.  This slice needs to be defined so that it avoids projecting
large amounts of foreground and background structures while still allowing enough numbers of neighbors
to obtain good statistics.  

In order to assess whether the alignments found in the previous subsection are
identifiable in observational catalogues analised this way, we reproduce the procedure mentioned
in the previous paragraph in our analysis of the numerical simulation which consists on projecting the 
structure onto the plane of the sky, assuming all objects within $\Delta v=750$km/s from the centre haloes 
to be exactly at the same distance from the observer.  

Therefore we only use information from two cartesian coordinates in the simulation, $x$ and $y$, 
and transform the third, $z$, into a velocity component by multiplying $z$ by the Hubble Constant 
and adding the $z$ component of the peculiar velocity. In order to complete the emulation of redshift space effects
suffered by observational surveys, we apply this transformation to the simulation particles as well 
as to the halo catalogue.  We then select "edge-on" haloes from the simulation, by requiring that 
their angular momentum is at an angle less than $60^o$ from the sky, represented by the $x-y$ plane
\footnote{
Such angle threshold is equivalent to 
a ratio of $0.5$ of the ellipsoidal image semi-axes for a perfect thin disc.}.
We can then study two projected correlation functions centred in these objects using DM 
particles, (i)
a correlation function calculated using pairs that subtend an angle
with the projected angular momentum of less than $30^o$, which we call
the parallel projected correlation function ($\omega_{||})$, and (ii) a perpendicular projected
correlation function ($\omega_{\bot}$) obtained by counting pairs that subtend an angle greater
than $60^o$ from the projected angular momentum.

Figure \ref{fig:projsim} shows the results from this procedure.  The left panels
show the results for low mass haloes in the small simulation (projected correlation
functions in the top-left, and ratios in the bottom-left), whereas the right panels
show the results for high mass haloes in the large simulation.
As can be seen, these results are qualitatively consistent with the alignment signal found in 
the previous subsection using the full three-dimensional correlation function, which indicates that 
projected correlation functions are a suitable tool to detect this effect.
The overall effect has a lower amplitude; high mass haloes show  a ratio between
structure in the directions perpendicular and parallel to their angular momenta of about a $30\%$
excess (for the 3D case, this was above $50\%$) and low mass haloes show a $10\%$ effect (compared to
$20-30\%$), which makes this detection a slightly more difficult task.

\section{Observations}

Galaxies with 
spiral and flattened spheroidal shapes projected onto the sky give a handle on their orientation 
with respect to the line of sight and, therefore, it is possible to
estimate in a statistical sense the direction of their angular momentum.
Spiral galaxies with round projected shapes are likely discs seen
face-on and therefore their angular momenta point in the direction perpendicular to the
plane of the sky; 
elongated projected shapes indicate discs seen edge-on, and therefore
their angular momenta point in the direction perpendicular to the long axis of the ellipsoid.
On the other hand, elliptical galaxies with round projected shapes can either
be nearly intrinsically spherical or flattened spheroids seen face-on;  however, elongated ellipticals
are more likely flattened spheroids seen edge-on since recent results on intrinsic elliptical shapes
indicate that these are more commonly oblate rather than prolate spheroids (e.g. Padilla \& Strauss,
2008).  Therefore, elongated ellipticals are likely to be also characterised by angular momenta pointing
in the direction of the plane of the sky roughly perpendicular to the major axis of the ellipsoid, 
as is the case for edge-on spiral galaxies.   These last two types of galaxies will be the
centre of the following analysis.
We notice that the amplitude of the angular momentum of flattened spheroidals is likely to
have a low amplitude relative to that of spiral galaxies.

The sample we selected to carry out our analysis is the spectroscopic SDSS-DR6,
which contains a total of $\simeq 580,000$ galaxies with spectroscopic redshifts and photometry
in five bands, $u,g,r,i,z$, as well as parameters determining the projected galaxy shapes consisting 
of the two semi-axes and the position angle of the ellipse that once convolved 
with the PSF (seeing) provides the best match
to the photometric image of each individual galaxy.  In the remainder of this paper, we only
consider SDSS-DR6 galaxies within the redshift range $0.02<z<0.09$.

\begin{table}
 \centering
 \begin{minipage}{80mm}
  \caption{Observational samples from the SDSS-DR6 spectroscopic main galaxy sample.
The first and second column indicate the sample name and number of galaxies in them,
the third column indicates the number density, and the fourth column the host-halo
mass.}
   \label{table}
  \begin{tabular}{@{}cccc@{}}
  \hline
  Sample & Members & $n/10^{-3}$h$^{-3}$Mpc$^3$ & Log$_{10}(M_{\rm host}/$h$^{-1}M_{\odot})$\\ 
  \hline
O1 & 132000&$10.4\pm0.8$&$12.50^{+0.30}_{-0.54}$\\
O2 & 59300 &$6.50\pm0.12$&$13.06^{+0.19}_{-0.27}$ \\
O3 & 72700 &$5.50\pm0.12$&$11.40^{+0.56}_{-0.80}$\\
O4 & 10900 &$7.15\pm0.17$&$12.19^{+0.45}_{-1.17}$\\
O5 & 1400  &$1.10\pm0.20$&$12.67^{+0.28}_{-0.48}$\\
  \hline
  \end{tabular}
  \end{minipage}
\end{table}

We study the alignment signal in the observational data for different subsamples defined by
galaxy colour and luminosity in order to detect trends that can be associated to the results
from the numerical simulation.  The subsamples we study are sample O1, consisting of
all the edge-on galaxies with axial ratio $b/a<0.7$ in the SDSS-DR6; 
O2, red galaxies with $g-r>0.7$; O3, blue galaxies with 
$g-r<0.7$; O4, faint galaxies with $M_r>-19.5$; and O5 composed by bright 
galaxies with $M_r<-19.5$.
The selection of subsamples is done so as to ensure a number of target galaxies to allow good
statistics ($> 1000$);  the number of galaxies in each sample is shown in Table \ref{table}, 
where we also show the galaxy number density calculated out to the redshift of completeness
of each sample.

In order to characterise in a quantitative way the subsamples selected from the SDSS-DR6,
we calculate the median mass of $\Lambda$CDM haloes predicted to show the same 2-halo regime
clustering amplitude
as the galaxies in each subsample.  We follow the procedure used in several previous works (cf.
Croft et al. 1999, Padilla et al., 2001) consisting on
calculating the auto-correlation function of galaxies in two coordinates, one parallel ($\sigma$) and one
perpendicular ($\pi$) to the line of sight, $\xi(\sigma,\pi)$, and integrating over the $\pi$ direction
to avoid the effect of redshift space distortions.  The result from this integration
is the projected correlation function (see for instance, Croft et al., 2001), $\Xi(\sigma)$,
\begin{equation}
\Xi(\sigma)=2\int_{\pi_{min}}^{\pi_{max}}\xi(\sigma,\pi) d\pi,
\label{eq:Xi}
\end{equation}
which in turn can be inverted to obtain
the real-space correlation function via
\begin{eqnarray}
&\xi(r)=\frac{-1}{\pi}\sum_{j\geq i}\frac{\Xi(\sigma_{j+1})-\Xi(\sigma_{j})}
{\sigma_{j+1}-\sigma_{j}}
 \ln{\left(\frac{\sigma_{j+1}+\sqrt{\sigma_{j+1}^{2}-
\sigma_{i}^{2}}}{\sigma_{j}+
\sqrt{\sigma_{j}^{2}-\sigma_{i}^{2}}}\right)},&
\end{eqnarray}
where the sum is performed over the bins in $\sigma$ where the projected correlation function
has been calculated.
The masses of host haloes can then be obtained by finding the bias between 
the resulting real-space correlation
functions and the matter correlation functions in a $\Lambda$CDM cosmology (both
in the 2-halo regime), which we obtain by Fourier
transforming the non-linear matter power spectrum (from Peacock \& Smith, 2000) for the
same cosmological parameters used in the numerical simulations.  With an estimate
of the bias, the mass of the host haloes can be found by using the Sheth, Mo \& Tormen (2001) formalism.
Note that this procedure is equivalent to measure the real-space halo-halo correlation function
for haloes of different masses (Sheth, Mo \& Tormen, 2001, Padilla \& Baugh, 2002), with the advantage
of avoiding large uncertainties from small samples of haloes restricted to narrow ranges of halo mass
in the simulation.
We show one example of this procedure in Figure \ref{fig:xir}, where we find the mass of haloes with
an equivalent clustering amplitude to the sample of red galaxies in the SDSS-DR6 (O2).  In this figure,
the top panel shows the auto-correlation function, $\xi(\sigma,\pi)$, of O2 galaxies, and the lower panel the
resulting real-space correlation function (symbols) and the best fit $\Lambda$CDM curve.  From this
analysis we find that red galaxies in the SDSS-DR6 (with $b/a<0.7$) are consistent with haloes
of a median mass of $\log_{10}(M/h^{-1}M_{\odot})=13.06^{+0.19}_{-0.27}$.  This provides us with
a quantitative way to compare the SDSS-DR6 subsamples to the halo samples in the numerical simulations.
Also, 
in order to infer this scale for our observational samples,
we will use the dependence of the 1- to 2-halo term transition on DM halo mass measured in our
the numerical simulations.

The other SDSS-DR6 galaxy samples are fitted by the correlation function of DM haloes of masses
indicated in Table \ref{table}.
Notice that our study covers over one and a half decades in the host-halo mass of SDSS-DR6 galaxies.

\begin{figure}
\begin{picture}(130,498)
\put(0,0){\psfig{file=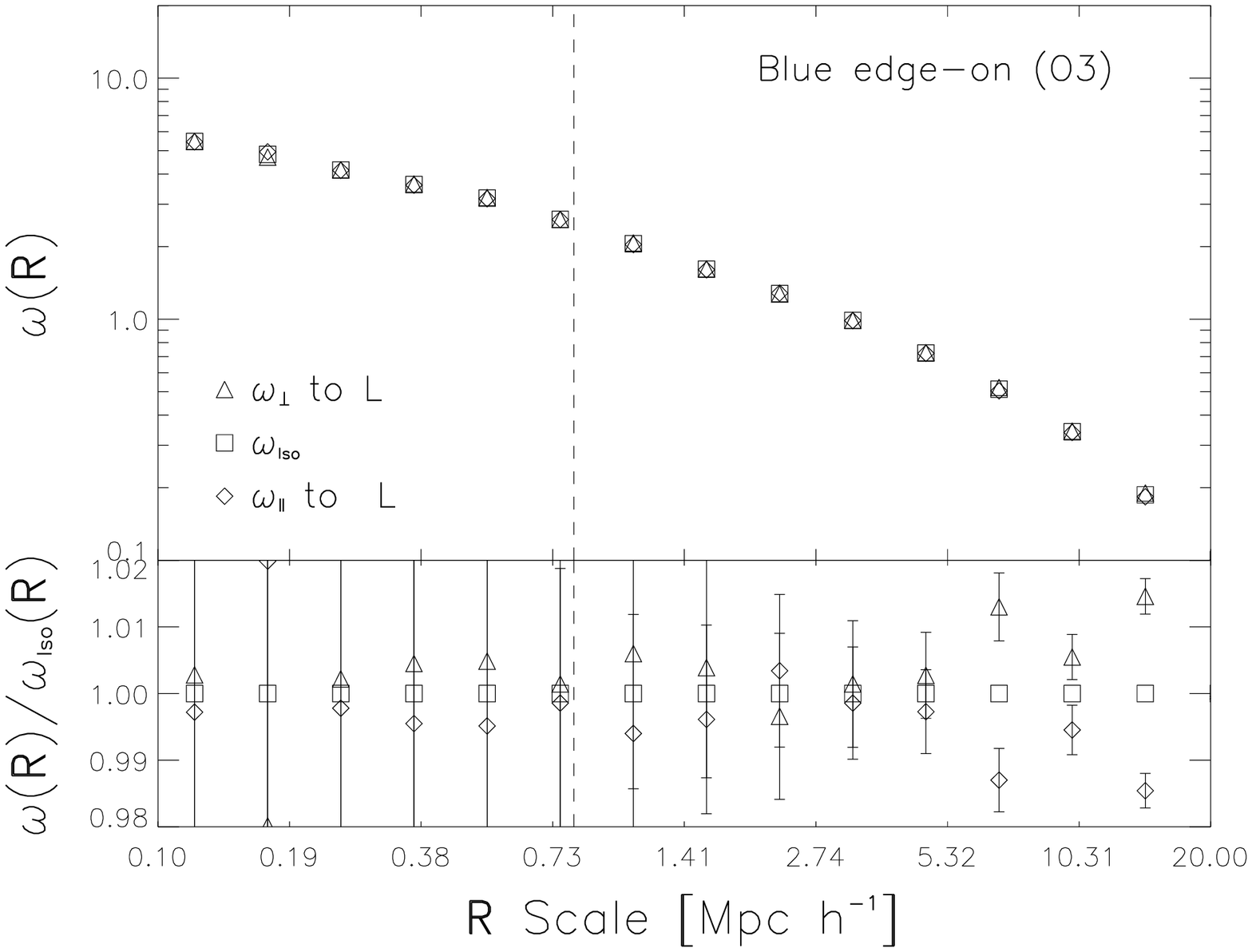,width=8.cm}}
\put(7,173){\psfig{file=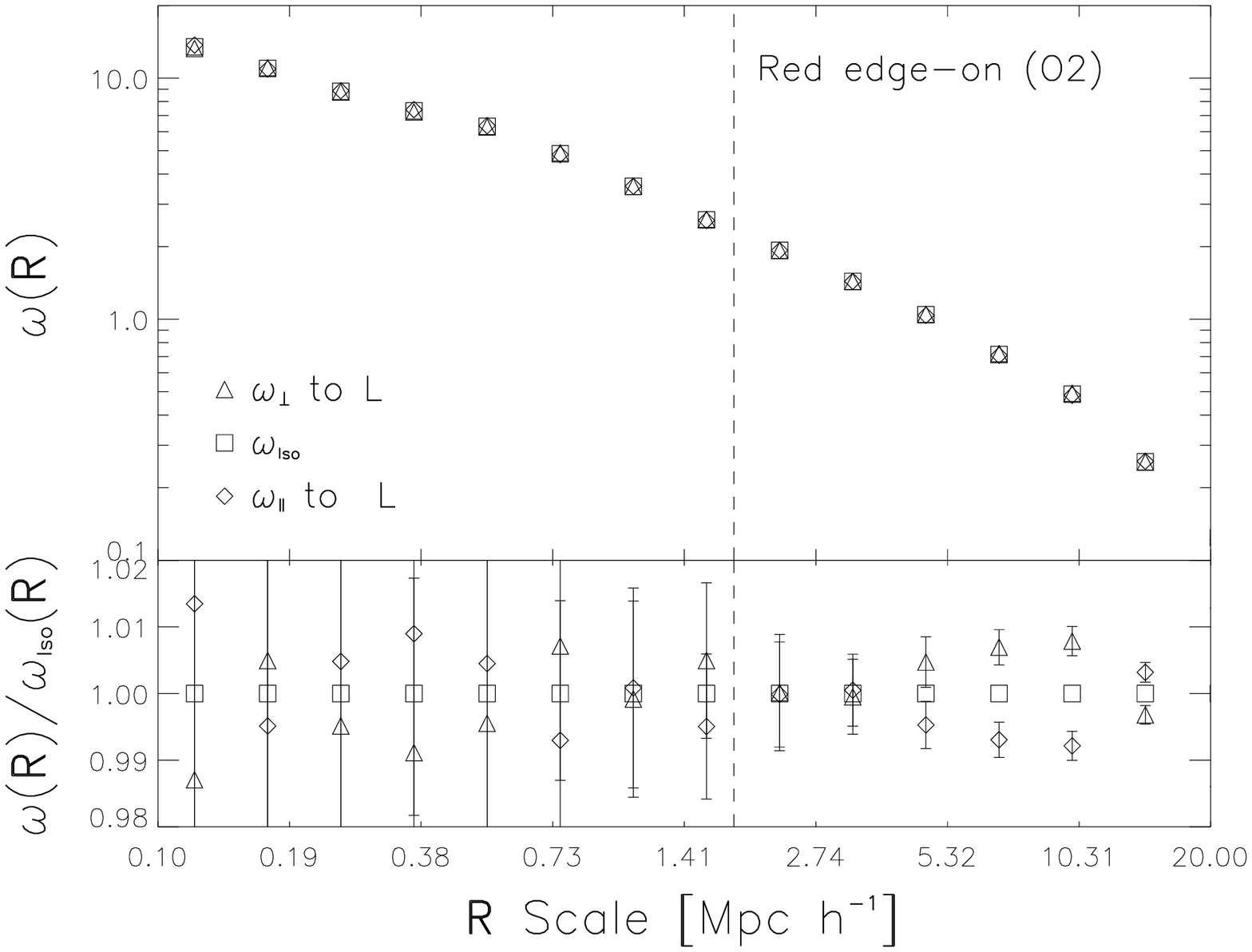,width=7.5cm}}
\put(7,340){\psfig{file=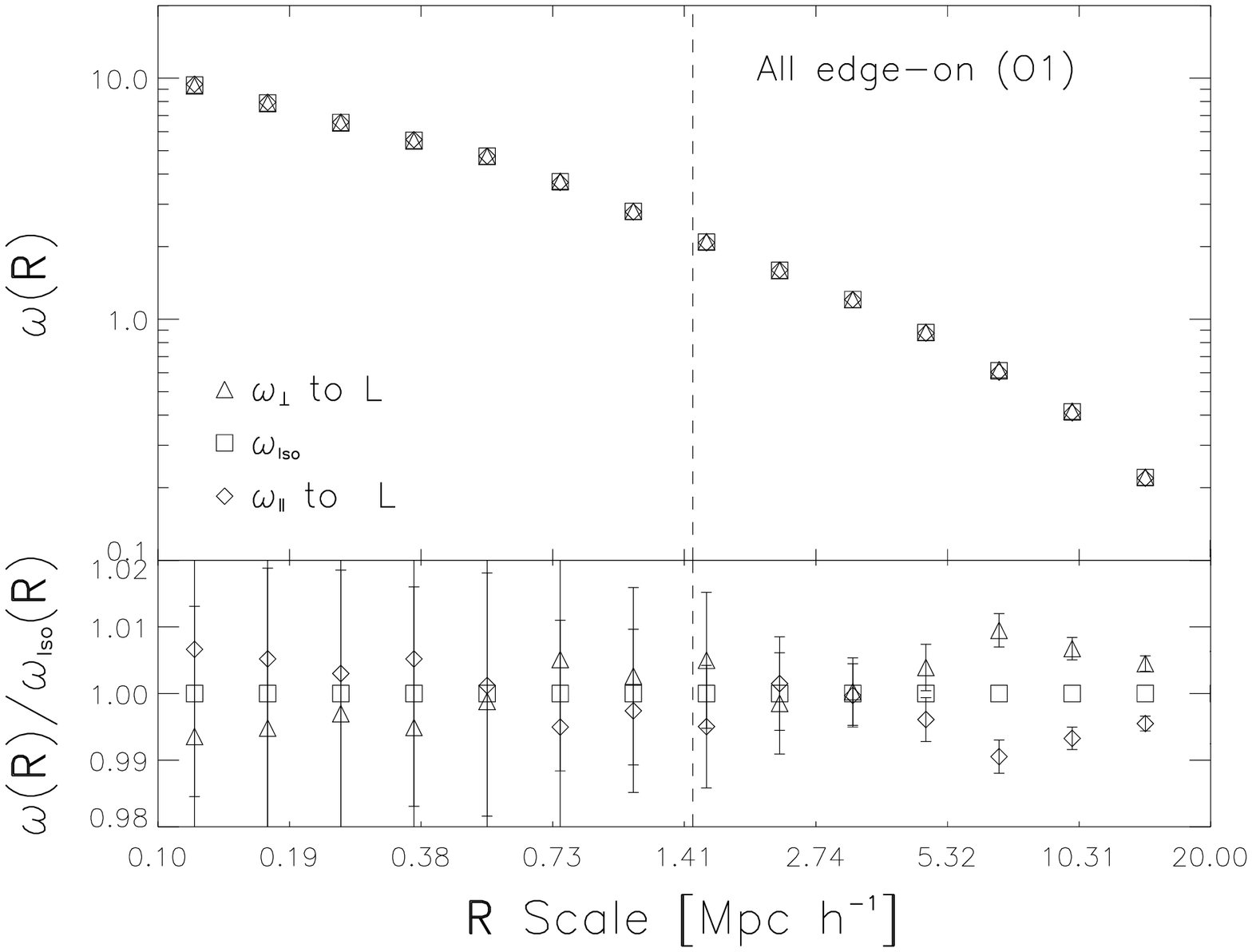,width=7.5cm}}
\end{picture}
       \caption{
Correlation functions around "edge-on" galaxies in the SDSS-DR6, in the direction
perpendicular and parallel to the angular momentum of centre haloes (triangles and
diamonds, respectively).  Top panel corresponds to sample O1 (full SDSS-DR6),
middle panel to O2 (red galaxies), and bottom panel to sample O3 (blue galaxies).
The vertical dashed lines show the transition from the 1- to the 2-halo regimes
corresponding to the host-halo mass obtained from the real-space correlation function
of the different samples.
}
       \label{fig:obsxi}
\end{figure}

As we demonstrated in the previous section both, the large scale structure and alignments surrounding galaxies, 
can be studied using the projected correlation function, $\omega(\sigma)$, measured in terms of
the projected separation, $\sigma$.  We now apply this method to our samples of edge-on
galaxies, and compare the outcome from using tracers in the direction parallel and 
perpendicular to the inferred angular momentum, to 
detect differences between the structure along these two directions.

The projected correlation functions are calculated using galaxies in a given subsample, 
cross-correlated to the full
spectroscopic SDSS-DR6 catalogue.  As in the 2-dimensional analysis carried out
in the numerical simulations, we calculate the cross-correlation for edge-on 
galaxies selected so that $b/a<0.7$.  Note that in the numerical simulation we
used an angle between the halo angular momentum and the line of sight of $60^o$, which
in the case of a perfect thin disk is equivalent to $b/a=0.5$, but is larger for
a thick disk as is the case of the intrinsic shapes of spiral galaxies or flattened ellipsoids
(Padilla \& Strauss, 2008).

Figure \ref{fig:obsxi} shows the measured correlation functions for Edge-on galaxies
for the three subsamples that show significantly different signatures of alignments with the
angular momentum.  These are the full sample of edge-on spirals, O1
(top panel),
the red galaxy sample, O2 (middle panel), and the blue galaxies, sample O3 (bottom panel).  
As in the results for the numerical simulation, triangles show the correlation with galaxies in the
direction of the angular momentum, diamonds in the perpendicular direction, and squares, when
using all neighbors.
A first glance at the shapes of these correlation functions indicates that there is not as clear
a transition between the 1- and 2-halo terms as in the results from the numerical simulations
(see Fig. \ref{fig:projsim}); in these samples the main effect is a transition to a roughly constant
slope for the $\log(\xi)$ vs. $\log \sigma$ relation.  
This may be an indication that the internal structure of haloes
is not traced in the same way by galaxies in the SDSS-DR6 and DM particles in the $\Lambda$CDM model.
Regarding the alignments, it can be seen that the three subsamples show the large-scale structure
preferentially aligned with the direction perpendicular to the angular momentum, specially for
the 2-halo regime, as was found in the cosmological simulation.
The difference between the two correlation functions is roughly
a $2-\sigma$ detection over the range of scales defined by $2<r/$h$^{-1}$Mpc$<20$ and corresponds to
about a $2-4\%$ effect.  

The analysis of the faint and bright samples (O4 and O5, respectively) provide alignment results in quantitative
agreement with the full galaxy sample, O1, with results within a $1-\sigma$ difference. 
Therefore, we will not show results from these two samples
in the remainder of this work.

A more quantitative comparison between observational and simulation results is presented in the following
section.

\section{Discussion}

In order to compare quantitatively the alignments found in the observations and in 
the numerical simulation, we use the transition scale from the 1- to the 2-halo regimes
inferred in the previous section to determine the degree of alignment found in
these two regimes separately. 
As in section \ref{simresult}, we estimate ratios between the correlation functions
in the directions perpendicular and parallel to the estimated galaxy angular momentum direction,
using all pairs within the 1- and 2-halo regimes.
This can then be directly compared to the results from the
numerical simulations for projected correlations.

\begin{figure}
\begin{picture}(180,440)
\put(0,0){\psfig{file=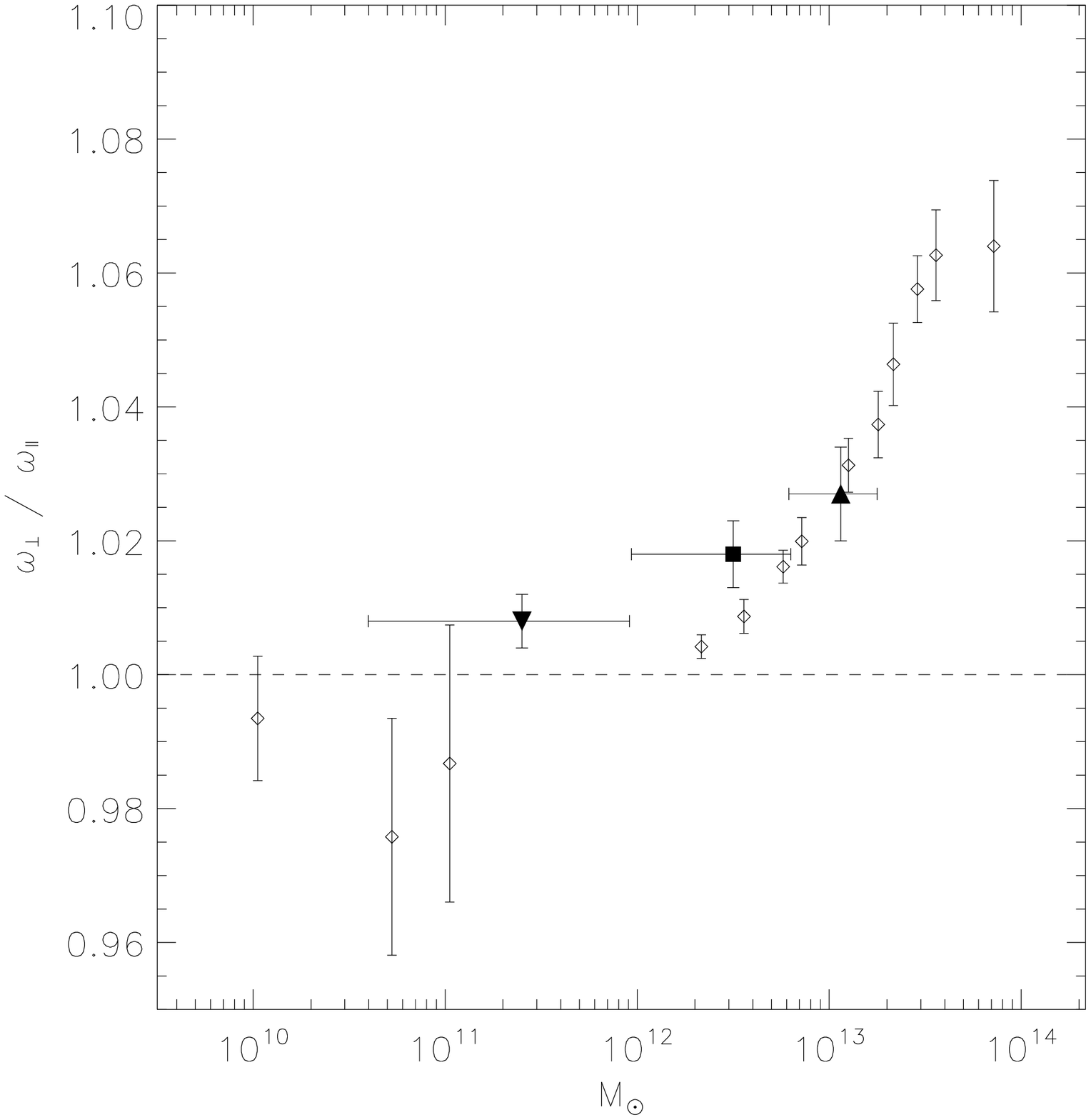,width=8.cm}}
\put(0,220){\psfig{file=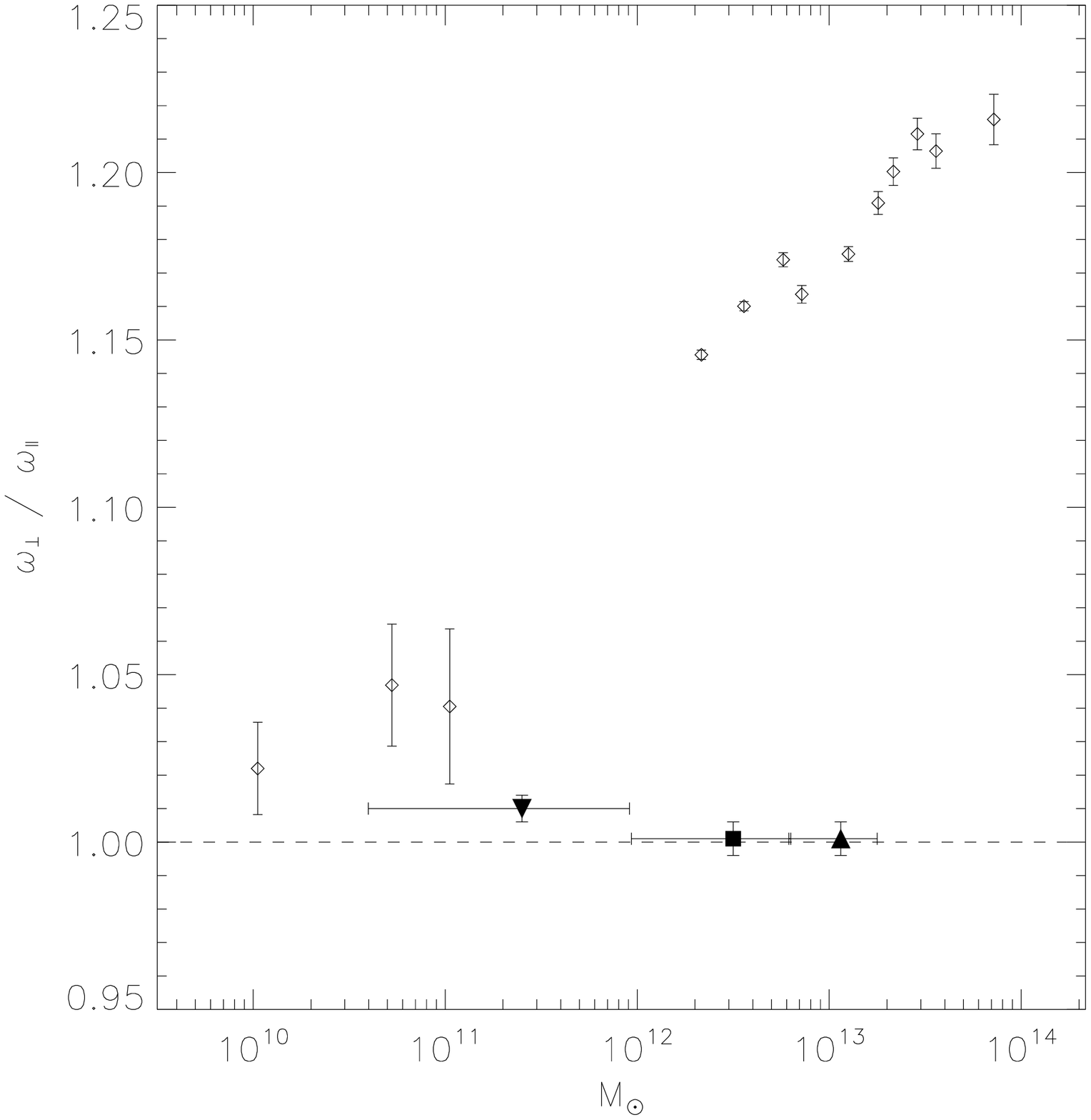,width=8.cm}}
\end{picture}
       \caption{Ratios between projected correlation functions in the 
                directions parallel and perpendicular to the projected angular momentum as 
		a function of halo mass in the numerical simulation (open diamonds). 
                Galaxy samples O1 (all galaxies), O2 (red galaxies), and 03 (blue galaxies), 
		are shown as filled symbols:
		squares, upward pointing triangles, and downward pointing triangles, respectively.
		Top panel: ratios for the 1-halo term. 
		Bottom panel: ratios for the 2-halo term. 
               }
       \label{fig:2drat}
\end{figure}

Figure \ref{fig:2drat} shows the ratios between projected correlation functions in the
directions parallel and perpendicular to the angular momentum as a function of
halo mass.  The results from the numerical simulations are shown as diamonds, and 
the results from the observational samples at the corresponding host-halo masses are shown as
filled symbols.
We show results for samples O1 (full SDSS-DR6, filled squares), O2 (red galaxies, filled upward pointing
triangles), and O3 (blue galaxies, filled downward pointing triangles).  The top panel
shows the results from the analysis of the 1-halo term, the lower panel from the 2-halo term.
  For the 2-halo term regime the resulting alignment signals for observational samples are
  $1.8\pm0.5\%$ for the full sample O1, $2.7\pm0.7\%$ for the red galaxy sample 02, and 
  $0.8\pm0.4\%$ for the blue galaxy sample 03.  
 The 1-halo term regime shows alignments of 
  $0.1\pm0.5\%$, $0.1\pm0.5\%$, and $1.0\pm0.4\%$ for samples O1, O2, and O3, respectively.
As can be seen, the alignment found in the 2-halo term is consistent between observational
samples and the numerical simulation (differences are at a maximum a $1-\sigma$
effect). 

 This agreement indicates that TTT in combination with hierarchical clustering are a 
viable theory since not only the angular momentum of
galaxies points in the direction perpendicular to the surrounding large-scale structure, but also the amplitude
of the effect is similar to what is found in the numerical simulations.  It should be noticed that this
would also indicate that the direction of the angular momentum in galaxies in these samples is tracing the
angular momentum of the host halo.  
The results for the 1-halo term, on the other hand, show that galaxies follow a different behaviour
than DM haloes as was suspected from the lack of a clear transition to the  1-halo term in the shape of the
SDSS projected correlation functions; at the range of masses corresponding to the observational samples, the
simulated haloes show very significant alignments, whereas the only sample of galaxies that a shows non-zero
alignment is that composed by blue galaxies (although still lower
than the simulation data, at a regime of low expected alignments).   
The combination of the 1- and 2-halo term results could indicate that the general galaxy population
in the SDSS-DR6 which our samples are cross-correlated to, 
can be used as good tracers of the large-scale structure (i.e. have not suffered important
changes to deviate from the angular momentum of their host haloes) but would fail to
serve as indicators of the internal distribution of matter in their host DM haloes; 
different astrophysical processes could produce this effect, specially for 
such low $M\simeq 10^{13}h^{-1}M_{\odot}$ halo masses\footnote{
There is observational evidence that groups of galaxies with masses above this limit
are characterised by intrinsic shapes consistent to those found in numerical simulations
\citep{Paz}; however it is difficult to directly measure the shapes of lower
mass systems due to discreteness effects.}.
This result is particularly interesting since blue galaxies do show
traces of an alignment in the same direction as the numerical simulation, which has a number
of possible explanations.  For instance,
the angular momentum of red galaxies could be subject to a reshuffle that 
adds enough noise to loose its alignment with the internal halo structure while
leaving the large-scale signal almost intact, an effect could also involve a positional reshuffle within 
the halo structure; another possibility is that red galaxies are
able to rearrange their companion galaxies inside DM haloes.
However, a more quantitative analysis is required in order to find the true underlying cause for this behaviour.
We acknowledge the possibility that observational biases not included in our analysis of
the numerical simulation are still affecting these measurements.

\section{Conclusions}

We study the alignments between the angular momentum of individual
objects and the large-scale structure in cosmological numerical simulations
and in the SDSS-DR6.  The angular momentum of DM haloes in $\Lambda$CDM simulations 
is found to be preferentially oriented in the direction perpendicular to the distribution
of matter in the range of separations corresponding to the 1- and 2-halo terms.  These
results are in agreement with the Tidal Torque Theory.  

We find that more massive haloes
show a higher degree of alignment, and that the 1-halo term shows a much higher alignment
than the 2-halo term.  We find a maximum 1-halo alignment of $\sim40\%$ for haloes
of $\simeq 10^{14}h^{-1}M_{\odot}$; at low masses, the 1-halo signal seems to tend to
a constant value of $20\%$. 
The 2-halo term alignment shows a maximum signal of $\simeq 15\%$ at high masses, and 
possible inversion at low masses where
the angular momentum might even preferentially point in the direction of the distribution
of mass on large scales.  The latter would be consistent with low mass groups
formed not too long ago from mass distributed in the surrounding areas
of the already-formed filaments or walls.

The alignment of the angular momentum with the large-scale structure found for low mass systems is in agreement
with previous works by \citep{Trujillo1} on redshift surveys, and by 
\citet{Brunino1,Patiri1,Cuesta1,Aragon1,Hahn1,Hahn2} on simulations. 
The main difference with these previous studies relies in that we quantify the strength of the 
alignment in terms of the two-point correlation function, as a function of halo mass. 
In particular, we find that a log-linear relation fits the dependence of alignment on mass, 
for both, the 1- and 2-halo term regimes. 
For the latter we observe a turning point between an angular momentum pointing in the direction
perpendicular to the structure, to no-alignment, and then to a regime where the angular momentum
is contained by the plane defined by the surrounding structure.  This occurs at a mass $\simeq M_*$,
in qualitative agreement with previous studies \citep{Bailin1,Hahn1,Hahn2}.
 
Combining our results for the 1- and 2-halo terms for haloes in the numerical simulation, namely
that the angular momentum of haloes is perpendicular to the bulk of the matter within haloes 
(1-halo term), and also to the surrounding large-scale structure (for masses $>M^*$),
our findings are in agreement with previous works who find that the major axis of haloes
is usually aligned with the large-scale structure
(Gottl{\"o}ber \& Turchaninov 2006, Bett et al. 2007,
Faltenbacher et al. 2002; Kasun \& Evrard 2005; Bailin 
\& Steinmetz 2005; Colberg, Krughoff \& Connolly 2005; Altay, Coldberg \& Croft 2006;
Basilakos et al. 2006; Ragone-Figueroa \& Plionis 2007, and references therein).
Regarding the 1-halo term signal, our results are also in qualitative agreement with
recent results by \citet{Knebe1} on the alignments of 
substructure within dark matter halos.

We use  numerical simulations to reproduce the procedure that can be applied to observational samples
using projected correlation functions.  We find that the alignment signal diminishes considerably
but that it can still be measured.

Therefore, we study edge-on galaxies in the SDSS-DR6,
and assign the direction perpendicular to the major semi-axis as the direction of the
angular momentum.  Since our samples are restricted to objects with flattened apparent
shapes ($b/a<0.7$), these include both edge-on spiral galaxies and flattened spheroids.  
We are able to detect alignments in all our galaxy samples. These include the full SDSS-DR6
edge-on galaxies, a sample composed by blue ($g-r<0.7$) galaxies, the sample of red ($g-r>0.7$)
galaxies, and a sample of faint and another composed by bright galaxies.  The latter 
two samples show a quantitative agreement with the results from the full sample of edge-on 
galaxies in the SDSS. 

In all cases we find a significant excess of structure
in the direction perpendicular to the angular momentum for the 2-halo term, in good agreement with the
numerical simulation results.  The 1-halo term results show either a null or lower alignment signal
than expected from the analysis of projected correlations in the numerical simulation.  
This would suggest the effects of astrophysical processes acting so that galaxies do not follow the DM
structure inside low mass haloes of $\leq 10^{13}h^{-1}M_{\odot}$ or lower masses, a result that 
complements previous studies of high mass groups in the SDSS with $M\gtrsim 10^{13}h^{-1}M_{\odot}$, 
where galaxies are found to follow the internal DM structure (e.g. Paz et al., 2006).  
This effect is apparently more
important for red, $g-r>0.7$ galaxies, indicating that there may be interesting astrophysical effects
at work at low mass dark-matter haloes which could produce changes in the orientation of galaxy angular momenta,
or the spatial re-distribution of galaxies within them.

\section*{Acknowledgments}
DJP acknowledges receipt of a fellowship of Consejo Nacional de Investigaciones Cientifico Tecnicas 
and the support of the European Union's ALFA-II programme, through LENAC, the Latin 
American European Network for Astrophysics and Cosmology. FAS acknowledges 
the receipt of an International Max-Planck Research School on Astrophysics fellowship.
NDP was supported by a Proyecto FONDECYT Regular 1071006.  This work was 
supported in part by Fondap "Center for Astrophysics" at Universidad Cat\'olica de Chile.
The authors are very grateful to Manuel Merch\'{a}n and Cintia Ragone, 
who kindly provided the numerical simulations and the 
"Friends of Friends" Code, respectively, used in this work. We acknowledge helpful
discussions with Diego Garcia Lambas, Manuel Merch\'{a}n and Ariel Sanchez. 
We are also grateful to the anonymous Referee, and to
Manolis Plionis, Bernardo Cervantes Sodi, Oliver Hahn and Sebastien Peirani, whose feedback 
helped us improve this work in a significant way. This research has made use of the 
NASA Astrophysics Data System.


\end{document}